**Super- and Hyperdeformed Isomeric States and Long-Lived Superheavy Elements***

A. Marinov[1], S. Gelberg[1], D. Kolb[2], R. Brandt[3], R. V. Gentry[4] and A. Pape[5]

[1]The Racah Institute of Physics, The Hebrew University, Jerusalem 91904, Israel
[2]Department of Physics, University GH Kassel, 34109 Kassel, Germany
[3]Kernchemie, Philipps University, 35041 Marburg, Germany
[4]Earth Science Associates, P O Box 12067, Knoxville, TN 37912-0067, USA
[5]IReS-UMR7500, IN2P3-CNRS/ULP, BP 28, F-67037 Strasbourg cedex 2, France

In recent years long-lived super- and hyperdeformed isomeric states have been discovered [1-4]. It was found that these isomeric states live much longer than the corresponding nuclei in their ground states (see Table 3 in [4]), and in addition, they have unusual radioactive decay properties. Thus, an isomeric state in the second minimum of the potential energy surface (a superdeformed (SD) isomeric state) may decay by relatively high energy and retarded α particles to the ground state, or to the normal deformed states, of the daughter nucleus, and also by low energy and enhanced α particles to the second minimum of the potential in the daughter. In addition it may also decay by very retarded proton radioactivity. An isomeric state in the third minimum of the potential (a hyperdeformed (HD) isomeric state) may decay by relatively high energy and retarded α particles to the second minimum of the potential in the daughter nucleus, or by low energy and enhanced α particles to the third minimum of the daughter. All these new and unusual radioactive decay properties have been found experimentally [1-4].

Based on these results the discovery [5,6] of element 112, back in 1971, produced via secondary reactions in CERN W targets irradiated with 24 GeV protons (see also [7,8]), has consistently been interpreted [4]. The long lifetime of several weeks, as compared to typical lifetimes of less than 1 ms [9], shows that a long-lived isomeric state rather than the normal ground state was produced in the reaction. The deduced fusion cross section in the region of a few mb, as compared to about 1 pb obtained in ordinary heavy ion reactions [9], is due to two effects:

a) The projectile in the secondary reaction experiments is not a normal nucleus in its ground state, but rather a fragment that has been produced by the high energy proton within about $2 \times 10^{-14}$ sec before interacting with another W nucleus in the target. During this short time it is at high excitation energy and quite deformed. Deformations increase the fusion cross section by several orders of magnitude as is well known from the sub-barrier fusion phenomenon [10] (see Fig. 10 in [8] and Fig. 7 in [4].)

b) The production of the compound nucleus in a super- or hyperdeformed isomeric state is much more probable than its production in the normal deformed ground state. The shapes of the compound nucleus in these isomeric states are close to those of the projectile-target combinations in their touching points. Therefore, much less inter-penetration and dissipation are needed in the formation of the compound nuclei in these isomeric states as compared to their production in the ground states (see Fig. 8 in [4]).

The discovery of the long-lived super- and hyperdeformed isomeric states enables one also to consistently interpret the unusually low energy and very enhanced α-particle groups seen in various actinide fractions separated from the same CERN W target. Thus the 5.14, 5.27 and 5.53 MeV α-particle groups, with corresponding half-lives of 3.8 ± 1.0 yr, 625 ± 84 d and 26 ± 7 d, seen in the Bk, Es and Lr-No sources, respectively, have consistently been interpreted, both from the point of view of their low energy and their five to seven orders of magnitude enhanced lifetimes, as possible $II^{min} \to II^{min}$, $III^{min} \to III^{min}$ and $III^{min} \to III^{min}$ transitions in $^{238}$Am, $^{247}$Es and $^{252}$No [4].

Based on the newly observed modes of radioactive decay of the super- and hyperdeformed isomeric states, consistent interpretations have recently been suggested by us for previously unexplained phenomena seen in nature [11,12]. These are the Po halos, the low-energy enhanced 4.5 MeV α-particle group proposed to be due to an isotope of a superheavy element with Z = 108, and the giant halos.

Po Halos were observed in mica minerals [13,14] where the concentric halos correspond to the decay chains of $^{210}$Po, $^{214}$Po and $^{218}$Po. Since the lifetimes of these isotopes are short, and halos belonging to their long-lived precursors from the $^{238}$U decay chain are absent, their origin is puzzling. It has been suggested [11,12] that their origin might be due to the existence of long-lived super- and hyperdeformed isomeric states in nuclei around $^{210}$Po, $^{214}$Po and $^{218}$Po which undergo β- and γ-decays to the ground states of these isotopes.[1]

The second unexplained phenomenon is the observation [16-19], in several minerals, of a low energy 4.5 MeV α-particle group with an estimated half-life of (2.5±0.5)x10$^8$ yr which, based on chemical behavior, has been suggested to be due to the decay of an isotope of Eka-Os (Z = 108; Hs). However, 4.5 MeV is a low energy compared to the predicted 9.5 - 6.7 MeV for β-stable isotopes of Hs [20-22], and $T_{1/2}$ = 2.5x10$^8$ yr is too short by a factor of 10$^8$, compared to predictions [23,24] from the lifetime versus energy relationship for normal 4.5 MeV α particles from Hs. It was recently shown [11,12], though, that these data can be quantitatively understood as a hyperdeformed to hyperdeformed transition from an isotope with Z = 108 and A ≈ 270. The low energy agrees with extrapolations from predictions [25] for $III^{min} \to III^{min}$ α transitions in the actinide region, and a half-life in the region of 10$^9$ yr is obtained if one takes into account in the penetrability calculations typical deformation parameters for a hyperdeformed nucleus.

Still another unexplained phenomenon is that of the giant halos [26]. Halos, with radii that fit the known ranges of 10 and 13 MeV α particles, have been seen in mica [26]. Unlike the situation with the Po halos, here it is not absolutely certain that their origin is from such high energy α particles [26-28]. However, if they are, then their existence is puzzling. For nuclei around the β-stability valley, 10 and 13 MeV α particles are respectively predicted [20-22] for Z values around 114 and 126. The estimated [23,24] half-life for 10 MeV α's in Z = 114 nuclei is about 1 sec, and for 13 MeV α's in Z = 126 nuclei, it is about 10$^{-4}$ sec. It is not clear how halos with such high-energy α particles and such short predicted lifetimes can exist in nature.
Here too an interpretation in terms of hyperdeformed isomeric states has been given [12]. A good candidate for the sequence of events producing the 10 MeV halo is a long-lived HD isomeric state decaying by a 4.8 MeV[2] $III^{min} \to III^{min}$ α transition, followed by β$^+$(EC) transitions to a normal state which decays by 10 MeV α particles. As a specific example, one may consider the following scenario where a HD isomeric state in $^{282}$114 decays by 4.8 MeV $\alpha$'s to a HD isomeric state in $^{278}$112, followed by two $\beta^+$(EC) decays to a normal deformed state or to the g.s. of $^{278}$110. This latter nucleus is predicted [20-22] to decay by 10 MeV $\alpha$ particles. For deformation parameters which are typical for a HD nucleus, the predicted $T_{1/2}$ value for a 4.8 MeV $III^{min} \to III^{min}$ $\alpha$ transition from $^{282}$114 is $10^8 - 10^{11}$ yr [12], and the sum of the two $Q_\beta$ values of above 6 MeV [20-22] makes the transition from the isomeric state in the third minimum to a normal state in one of the daughter nuclei possible.

---

[1] However, see Ref. [15].
[2] As a matter of fact a halo with a radius which may correspond to 4.8 MeV α particles is also seen in the 10-MeV halo [12].

Similarly, for the 13 MeV halo a possible scenario has been suggested [12] where a HD isomeric state in $^{316}$126 decays by a $III^{min} \to III^{min}$ low-energy $\alpha$ transition of about 5.1 MeV[3] to $^{312}$124, followed by two $\beta^+$(EC) transitions, leading to the g.s. of $^{312}$122. The $^{312}$122 nucleus is predicted [20,21] to decay by $\alpha$ particles of around 13 MeV. For a 5.1 MeV HD to HD $\alpha$ transition from $^{316}$126, the predicted [12] half-life, using the parameters of Ref. [29] for a HD shape of $^{232}$Th, is $3 \times 10^{11}$ yr. (Larger deformation parameters give shorter lifetimes).

The above analysis suggests that primordial heavy and superheavy nuclei in long-lived isomeric states might exist in nature. A program has been started to search for such nuclei in petzite ($Ag_3AuTe_2$) and monazite ($(Ce,La,Th)PO_4$) minerals using the accelerator mass spectrometry (AMS) system [30,31] of the Weizmann Koffler Pelletron accelerator in Rehovot. The first mineral was chosen since there is an indication that induced Po X-rays in such a mineral from Romania has been observed [32]. The second mineral is the same as the one where the giant halos were found [26]. A progress report will be given.

---

[3] As a matter of fact a halo with a radius which may correspond to 5.1 MeV $\alpha$ particles is also seen in the 13-MeV halo [12].